\documentclass[prb,twocolumn,showpacs,preprintnumbers,amsmath,amssymb,superscriptaddress]{revtex4}
\usepackage{graphicx}
\usepackage{dcolumn}
\usepackage{bm}

\begin{document}
\title{Magneto-transmission of multi-layer epitaxial graphene and bulk graphite: \\
A comparison}
\author{M. Orlita}
\email{orlita@karlov.mff.cuni.cz} \affiliation{Grenoble High Magnetic Field
Laboratory, CNRS, BP 166, F-38042 Grenoble Cedex 09, France}
\affiliation{Institute of Physics, Charles University, Ke Karlovu 5, CZ-121~16
Praha 2, Czech Republic} \affiliation{Institute of Physics, v.v.i., ASCR,
Cukrovarnick\'{a} 10, CZ-162 53 Praha 6, Czech Republic}
\author{C. Faugeras}
\affiliation{Grenoble High Magnetic Field Laboratory, CNRS, BP 166, F-38042
Grenoble Cedex 09, France}
\author{G. Martinez}
\affiliation{Grenoble High Magnetic Field Laboratory, CNRS, BP 166, F-38042
Grenoble Cedex 09, France}
\author{D. K. Maude}
\affiliation{Grenoble High Magnetic Field Laboratory, CNRS, BP 166, F-38042
Grenoble Cedex 09, France}
\author{J. M. Schneider}
\affiliation{Grenoble High Magnetic Field Laboratory, CNRS, BP 166, F-38042
Grenoble Cedex 09, France}
\author{M. Sprinkle}
\affiliation{School of Physics, Georgia Institute of Technology, Atlanta,
Georgia 30332, USA} \affiliation{Institut N\'{e}el/CNRS-UJF BP 166,
F-38042 Grenoble Cedex 9, France}
\author{C. Berger}
\affiliation{School of Physics, Georgia Institute of Technology, Atlanta,
Georgia 30332, USA} \affiliation{Institut N\'{e}el/CNRS-UJF BP 166,
F-38042 Grenoble Cedex 9, France}
\author{W. A. de Heer}
\affiliation{School of Physics, Georgia Institute of Technology, Atlanta,
Georgia 30332, USA}
\author{M. Potemski}
\affiliation{Grenoble High Magnetic Field Laboratory, CNRS, BP 166, F-38042
Grenoble Cedex 09, France}
\date{\today}

\begin{abstract}
Magneto-transmission of a thin layer of bulk graphite is compared
with spectra taken on multi-layer epitaxial graphene prepared by
thermal decomposition of a SiC crystal. We focus on the spectral
features evolving as $\sqrt{B}$, which are evidence for the presence
of Dirac fermions in both materials. Whereas the results on
multi-layer epitaxial graphene can be interpreted within the model
of 2D Dirac fermions, the data obtained on bulk graphite can only be
explained taking into account the 3D nature of graphite, e.g. by
using the standard Slonczewski-Weiss-McClure model.
\end{abstract}


\maketitle

\section{Introduction}

Within the last few years, the properties of massless and massive
``relativistic'' particles have been probed in magneto-optical
experiments carried out on several carbon-based systems: graphene
monolayer\cite{JiangPRL07,DeaconPRB07} and
bilayer,\cite{HenriksenPRL08} multi-layer epitaxial
graphene,\cite{SadowskiPRL06,SadowskiSSC07,PlochockaPRL08,OrlitaCM08II}
as well as bulk
graphite.\cite{LiPRB06,OrlitaPRL08,OrlitaJPCM08,Garcia-FloresCM08}
These experiments offer an important  insight into the electronic
structure of these materials, in particular, they directly
demonstrate the unusual $\sqrt{B}$-scaling of Landau levels (LLs) of
Dirac fermions. Naturally, these experiments stimulated intensive
theoretical work in this
area.\cite{IyengarPRB07,GusyninPRL07,AbergelPRB07,KoshinoPRB08,BychkovPRB08}

Recently, remarkable experiments have been performed on the same
systems without applied magnetic field. A universal value of optical
conductivity was demonstrated in graphene\cite{NairScience08} as
well as in bulk graphite.\cite{KuzmenkoPRL08} The optical response
of graphene monolayer and bilayer samples as a function of the gate
voltage was reported in several experimental works, see
Refs.~\onlinecite{WangScience08,LiNaturePhys08,KuzmenkoCM08,ZhangCM08},
and this topic was addressed also
theoretically.\cite{McCannSSC07,StauberPRB08,NicolPRB08} First
results of time-resolved optical experiments on multi-layer graphene
are also available.\cite{DawlatyAPL08,SunPRL08}

In this paper, we present a direct comparison of the
magneto-transmission spectra taken on a thin layer of bulk graphite
and multi-layer epitaxial graphene. We focus on spectral features
exhibiting a $\sqrt{B}$-dependence, which corresponds to the optical
response of Dirac fermions, and show that there are essential
differences in the spectra of both materials.

\section{Experimental details}

The investigated multi-layer graphene sample was prepared by thermal
decomposition on the carbon face of a 4H-SiC
substrate\cite{BergerJPCB04,BergerScience06} and contains around
$\sim$100 graphene layers. The thin layer of bulk graphite was
prepared by simple exfoliation as described in
Ref.~\onlinecite{OrlitaPRL08}. As no significant difference in FIR optical
response of natural graphite and HOPG was found,\cite{OrlitaJPCM08}
we present only results taken on natural graphite due to its higher
crystalline quality. Both samples were characterized using
micro-Raman. In the case of multi-layer epitaxial graphene, the
Raman signature of decoupled layers, equivalent to a single flake of
exfoliated graphene,\cite{FerrariPRL06} as well as of additional
graphite residuals were found, see Ref.~\onlinecite{FaugerasAPL08}. The
Raman spectra taken on a thin graphite layer dominantly showed  a
multi-component 2D band typical of many Bernal-stacked sheets, but
some minor traces of decoupled layers were found, see the discussion
in Ref.~\onlinecite{OrlitaJPCM08}.

To measure the FIR transmittance of the sample, the radiation of globar, delivered via
light-pipe optics to the sample and detected by a Si bolometer
placed directly below the sample, was analyzed by a Fourier transform spectrometer. All
measurements were performed in the Faraday configuration with the
magnetic field applied normal to graphene/graphite layers. All the
spectra were taken with non-polarized light in the spectral range of
5-350~meV, limited further by several regions of low tape
transmissivity or the SiC opacity, see grey areas in Fig.~\ref{Comparison}.

\section{Results and Discussion}

\begin{figure}[t]
\scalebox{1.8}{\includegraphics*[10pt,10pt][138pt,253pt]{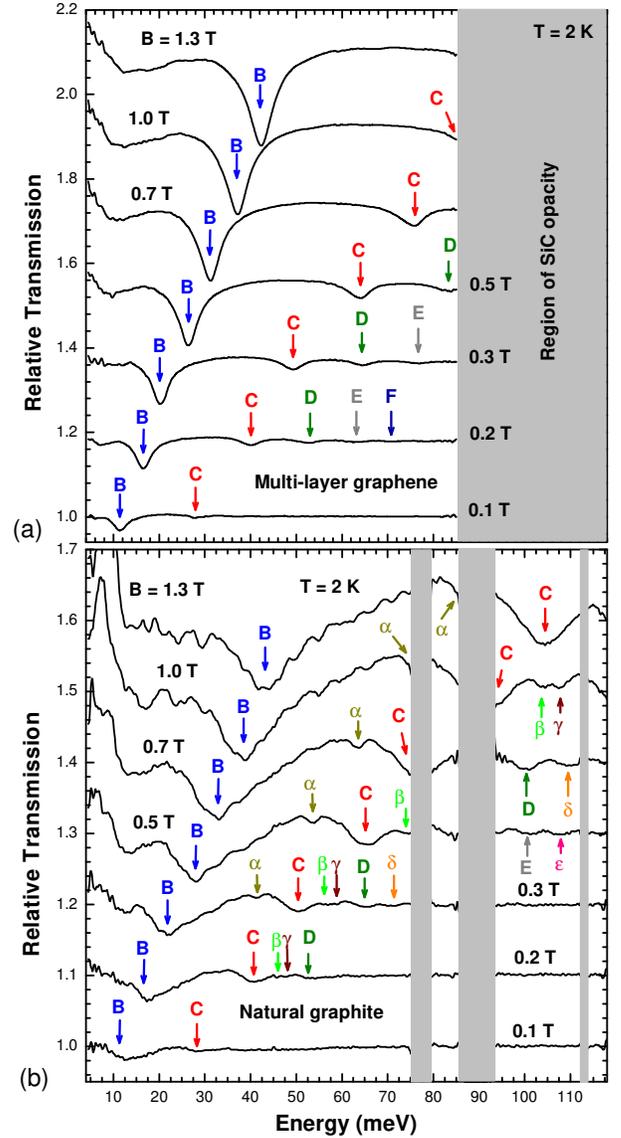}}
\caption{\label{Comparison}  Transmissions spectra of multi-layer
epitaxial graphene (a) and  bulk graphite (b) for selected magnetic
fields at $T=2$~K. The absorption lines corresponding to
dipole-allowed transitions in graphene are denoted by Roman letters.
Greek letters are used for additional transitions which scale as
$\sqrt{B}$ and are only found in spectra taken on bulk graphite. For
clarity, the spectra in part (a) and (b) were shifted by 0.18 and
0.10, respectively.}
\end{figure}

A comparison of the transmission spectra taken on multi-layer
epitaxial graphene and a thin layer of bulk graphite is presented in
Fig.~\ref{Comparison}. Starting with results taken on multi-layer
layer graphene in Fig.~\ref{Comparison}a, a series of absorption
lines is observed and denoted by Roman letters, following the
notation introduced in Ref.~\onlinecite{SadowskiPRL06}. Assuming the LL
spectrum of graphene, $E_n=\tilde{c}\sqrt{2e\hbar B n}$, together
with the corresponding selection rules for dipole-allowed
transitions $\Delta n=\pm1$, the observed absorption lines B,C,D,E
and F can be clearly identified as inter-LL transitions
L$_{-m}\rightarrow$L$_{m+1}$ and L$_{-(m+1)}\rightarrow$L$_m$ with
$m=0,1,2,3$ and 4, respectively. The Fermi velocity is found to be
$\tilde{c}=(1.02\pm0.01)\times10^{6}$~m.s$^{-1}$. Hence, these
results are fully consistent with a model of 2D Dirac fermions and
individual sheets in multi-layer epitaxial graphene indeed behave as
if they are electronically decoupled. This was recently explained by
a random mutual rotation of adjacent sheets in this
material.\cite{HassPRL08} Similar spectra were also measured on a
single sheet of exfoliated graphene,\cite{JiangPRL07} however, the
Fermi velocity seems to be enhanced by amount of~$\approx$10\% and
these data also suggest some influence of electron-electron
interaction, not observed in transmission spectra of multi-layer
epitaxial graphene.

The results obtained on a thin layer of bulk graphite, see Fig.~1b,
are more complex. Basically, all the absorption lines observed in
multi-layer graphene are also found in spectra of bulk graphite with
practically the same Fermi velocity
$\tilde{c}=(1.02\pm0.02)\times10^{6}$~m.s$^{-1}$. This justifies the
same notation using the Roman letters. In addition to these lines,
another series of transitions, denoted by Greek letters and clearly
exhibiting the $\sqrt{B}$-dependence, is present in transmission
spectra. These lines cannot be assigned to any dipole-allowed
transitions between LLs in graphene. Nevertheless, the energies of
the additional lines $\alpha,\gamma,\delta$ and $\varepsilon$
exactly match to transitions symmetric around the Dirac point,
L$_{-m}\rightarrow$L$_{m}$, with indices $m=1,2,3$ and 4,
respectively. The $\beta$ line can be identified as transitions
L$_{-1(-3)}\rightarrow$L$_{3(1)}$.

\begin{figure*}
\scalebox{0.75}{\includegraphics*[8pt,409pt][587pt,713pt]{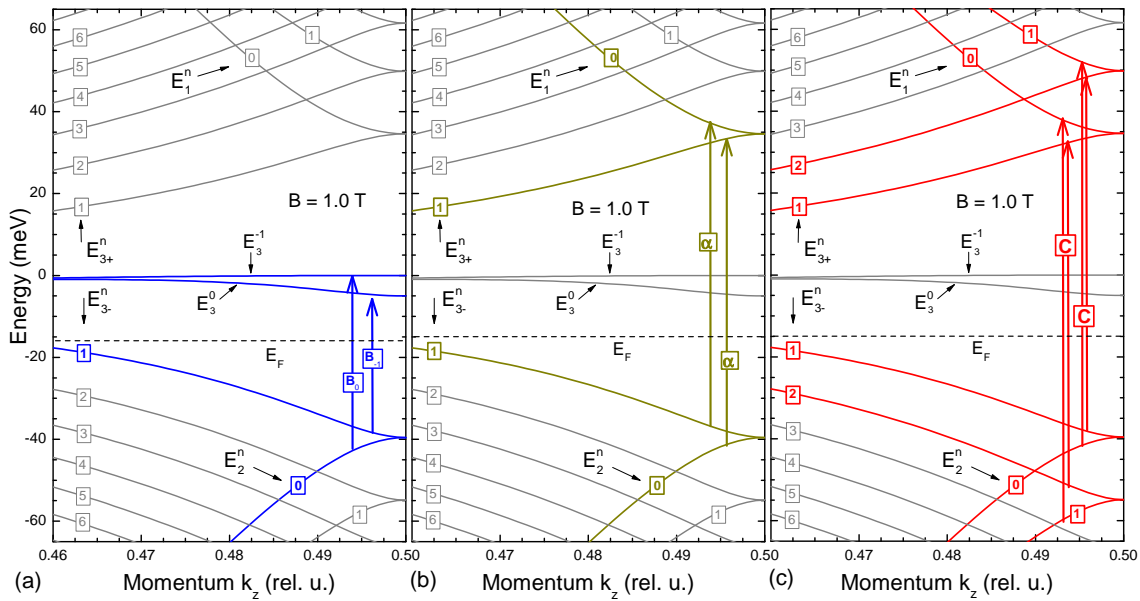}}
\caption{\label{Asymmetry} LL structure of bulk graphite in the
vicinity of the $H$ point. Dipole-allowed transitions between LLs
depicted in blue, dark yellow  and red colors in part (a), (b) and
(c) contribute to the B,$\alpha$ and C lines in the transmission
spectrum, respectively, cf. Fig.~\ref{Comparison}b. The parameters
$\Delta=-5$~meV and $\tilde{c}=1.02\times10^6$~m.s$^{-1}$ are used
for this calculation. The effects of trigonal warping are
neglected.}
\end{figure*}

To explain the absorption lines denoted by the Greek letters we have
to abandon the simplified model of 2D Dirac fermions and consider
the full band structure of bulk graphite. According to the standard
SWM model,\cite{SlonczewskiPR58,McClurePR57} Dirac fermions are
located in the vicinity of the $H$ point, where the bands $E_1$,
$E_2$ and the doubly degenerate $E_3$ are close to the Fermi level.
If the magnetic field is applied along the $c$-axis of the crystal,
Landau levels or more accurately, Landau bands are created, see
Fig.~\ref{Asymmetry}, having at the $H$ point ($k_z=0.5$)  an
analytical form\cite{McClurePR60,ToyPRB77} ($n\geq 1$):
\begin{eqnarray}\label{EnergyLL}
E_3^{-1}& = & 0 \nonumber \\
E_3^0 &= &\Delta \nonumber \\
E_{3\pm}^{n} &=&E_{1,2}^{n-1}=\frac{\Delta}{2}\pm\sqrt{\frac{\Delta^2}{4}+\xi B n},
\end{eqnarray}
where $\Delta$ denotes the pseudogap, i.e. the distance of $E_1$ (and also
$E_2$) and $E_3$ bands at $k_z=0.5$ and $B=0$. The parameter $\xi$ is related to the
Fermi velocity as $\xi=2\tilde{c}^2e\hbar$.

The form of LLs~\eqref{EnergyLL} implies several important $H$
point-related optical properties of graphite. It suggests that the
LL energy spectrum typical of graphene is present also at the $H$
point in graphite, when the pseudogap $|\Delta|$ is small in
comparison to the energies of the LLs. Experimentally, $|\Delta|$ is
found to be significantly below 10~meV, see
Refs.~\onlinecite{ToyPRB77,OrlitaPRL08,GruneisCM08}, and therefore, the
magnetic fields above 100~mT are sufficient to insure this
condition. On the other hand, each LL in Eq.~\eqref{EnergyLL} is
(with the exception of $E_3^0$ and $E_3^{-1}$) doubly degenerate,
$E_{3+}^{n}=E_{1}^{n-1}$ and $E_{3-}^{n}=E_{2}^{n-1}$, see
Fig.~\ref{Asymmetry}. This double degeneracy is in addition to the
spin and valley degeneracies present in graphene. Obviously, taking
account of the same selection rules $\Delta n=\pm1$, we obtain a
considerably richer set of possible dipole-allowed transitions for
graphite in comparison to graphene. For instance, the transition
L$_{-1}\rightarrow$L$_{1}$ is strictly forbidden in graphene,
nevertheless the absorption line at this energy ($\alpha$
transition) is observed in bulk graphite at the $H$ point due to the
dipole-allowed transitions $E_{3-}^{1} \rightarrow E_{1}^{0}$ and
$E_{2}^{0} \rightarrow E_{3+}^{1}$, see Fig.~\ref{Asymmetry}b.
Similarly, the lines $\gamma,\delta$ and $\varepsilon$ are detected
due to transitions $E_{3-}^{m+1} \rightarrow E_{1}^{m}$ and
$E_{2}^{m} \rightarrow E_{3+}^{m+1}$ for $m=1,2$ and 3,
respectively. The remaining $\beta$ line can be identified as
$E_{3-}^{1} \rightarrow E_{1}^{2}$ and $E_{2}^{2} \rightarrow
E_{3+}^{1}$. Hence, the presence of absorption lines denoted by
Greek letters is qualitatively consistent with the standard SWM
model of graphite. Nevertheless, we are aware of inconsistency of
our results with predictions of Koshino and
Ando.\cite{KoshinoPRB08} They expect that for graphite layer with
several tens or more Bernal-stacked sheets, the intensities of
$\sqrt{B}$-dependent absorption lines should be negligible in
comparison with spectral features arising in massive electrons
around the  $K$ point, which evolve nearly linearly with $B$.

Another difference of spectra Figs.~\ref{Comparison}a and b is an
apparent asymmetry of the B line in bulk graphite at low magnetic
fields, whereas the equivalent line in the spectra of multi-layer
graphene remains perfectly symmetric. To explain this difference, we
must extend our considerations to the near vicinity of the $H$ point
and take into account the $k_z$-dependence of LLs. Such LL structure
calculated within the standard SWM model is shown in
Fig.~\ref{Asymmetry}a, where the two components, B$_{0}$ and
B$_{-1}$ of which the B line consists, are denoted by arrows. Note
that the relatively low value of the pseudogap $|\Delta|$ does not
allow to directly resolve the splitting of the B line into these
components. At low magnetic field, when the Landau band $E_{3-}^1$
is nearly parallel with $E_{3}^{0}$ and/or when it is partially or
completely located above the Fermi level, we can expect rather
asymmetrical shape of the B line with a well-pronounced high energy
tail due to the transition $E_{2}^{0} \rightarrow E_{3}^{-1}$. With
increasing magnetic field, this asymmetry should gradually
disappear, as indeed observed experimentally,
see~Fig.~\ref{Comparison}b. From the lineshape, we roughly estimate
the position of the Fermi level slightly below $\approx$-20~meV, but
the situation is here complicated by the fact that also decoupled
graphene layers in graphite, revealed by the Raman spectroscopy or
detected in scanning tunneling spectroscopy
experiments,\cite{LiCM08} can contribute to the final shape of the
B line.

Note, that for the sample of multi-layer graphene investigated here,
the fully symmetrical B line survives in transmission spectrum down
to magnetic fields of 40~mT, when the Fermi level is located at
$\approx$7~meV from the Dirac point.\cite{OrlitaCM08II} The
presented transmission experiment thus probes mainly undoped
graphene sheets ($n_0\approx 5\times10^{9}$~cm$^{-2}$), which are
located further from the SiC substrate. The layer(s) in the
immediate vicinity of the substrate are highly conductive (above
$n_0\approx 10^{12}$~cm$^{-2}$) and can be probed in transport
experiments.\cite{BergerJPCB04,BergerScience06}

Analogous considerations based on the shapes of Landau bands in the
vicinity of the H point, can also qualitatively explain the
significant differences in widths of individual absorption lines.
For instance, the C line is roughly three times broader in
comparison to the $\alpha$ line, see Fig.~\ref{Comparison}b, in
spite of the fact that they both originate in the same LLs. For a
simple explanation of this fact, we just need to realize that the
$\alpha$ line is composed of transitions $E_{3-}^{1} \rightarrow
E_{1}^{0}$ and $E_{2}^{0} \rightarrow E_{3+}^{1}$. If we take
account of the near vicinity of $k_z=0$, we find out that both pairs
of these LLs are nearly parallel, see Fig.~\ref{Asymmetry}b, and
give thus a rather sharp $\alpha$ line in comparison to the C line.
This line is composed of transitions $E^{1(2)}_{3-}\rightarrow
E_{3+}^{2(1)}$ and $E^{1(2)}_{2}\rightarrow E_{1}^{2(1)}$, see
Fig.~\ref{Asymmetry}c, which occur between pairs of LLs having their
derivatives of opposite signs, which causes the broadening of the C
line. This explanation can be straightforwardly generalized for all
transitions, giving qualitative explanation of the differences in
linewidths simply using the standard SWM model.

In principle, another and even more appealing interpretation of our
data could assume that we actually see the response of decoupled
graphene layers instead of the $H$ point of bulk graphite. The
presence of additional ``Greek'' lines in spectra could be then a
consequence of some perturbation of the graphene system leading to a
relaxation of the selection rules $\Delta n=\pm1$. Even though the
appearance of some decoupled graphene sheets in our sample is very
likely, we have the following arguments to support our ``bulk''
interpretation: i) Our experiments\cite{OrlitaPRL08,OrlitaJPCM08}
were performed on several samples prepared from two different types
of bulk graphite and the almost identical results, concerning the
presence, position, lineshape and mainly mutual intensities of all
the $\sqrt{B}$-scaled absorption lines were obtained. ii) The
``Greek'' transitions show significantly smaller linewidth in
comparison to ``Roman'' ones, e.g. compare lines $\alpha$ and C,
which is unlikely to be caused by any perturbation relaxing the
selection rules in graphene.

In both materials, spectral features evolving nearly linearly with
$B$ are also observed, see low energy part of spectra in
Figs.~\ref{Comparison}a and b, and represent thus an evidence for
massive particles. In bulk graphite, can we relate these transitions
to the $K$ point, i.e. to massive electrons, as discussed already in
Ref.~\onlinecite{OrlitaPRL08}. The optical response of massive electrons
in the magnetic field was thoroughly analyzed e.g. in
Refs.~\onlinecite{ToyPRB77,LiPRB06}. In case of multi-layer graphene, the
spectral features linear in $B$ cannot be explained unless some
part of the sample is created from bulk graphite or at least from
few-layer graphene stacks. Indeed, the traces of Bernal-stacked
layers were found in micro-Raman experiment, as mentioned above.

\section{Conclusions}

We have compared the optical response of Dirac fermions in bulk
graphite and multi-layer epitaxial graphene in low magnetic fields.
Whereas the results obtained on multi-layer graphene fully
correspond to expectations for dipole-allowed transitions in a 2D
gas of Dirac particles, the transmission spectra taken on bulk
graphite appear to be more complex. The standard
Slonczewski-Weiss-McClure model is found, in the latter case to be
sufficient to explain the existence of all absorption lines scaling
as $\sqrt{B}$ as well as their individual lineshapes.

\begin{acknowledgments}
The present work was supported by the European Commission through
Grant No. RITA-CT-2003-505474, by contract ANR-06-NANO-019 and projects
MSM0021620834 and KAN400100652.
\end{acknowledgments}


\begin{thebibliography}{38}
\expandafter\ifx\csname natexlab\endcsname\relax\def\natexlab#1{#1}\fi
\expandafter\ifx\csname bibnamefont\endcsname\relax
  \def\bibnamefont#1{#1}\fi
\expandafter\ifx\csname bibfnamefont\endcsname\relax
  \def\bibfnamefont#1{#1}\fi
\expandafter\ifx\csname citenamefont\endcsname\relax
  \def\citenamefont#1{#1}\fi
\expandafter\ifx\csname url\endcsname\relax
  \def\url#1{\texttt{#1}}\fi
\expandafter\ifx\csname urlprefix\endcsname\relax\def\urlprefix{URL }\fi
\providecommand{\bibinfo}[2]{#2}
\providecommand{\eprint}[2][]{\url{#2}}

\bibitem[{\citenamefont{Jiang et~al.}(2007)\citenamefont{Jiang, Henriksen,
  Tung, Wang, Schwartz, Han, Kim, and Stormer}}]{JiangPRL07}
\bibinfo{author}{\bibfnamefont{Z.}~\bibnamefont{Jiang}},
  \bibinfo{author}{\bibfnamefont{E.~A.} \bibnamefont{Henriksen}},
  \bibinfo{author}{\bibfnamefont{L.~C.} \bibnamefont{Tung}},
  \bibinfo{author}{\bibfnamefont{Y.-J.} \bibnamefont{Wang}},
  \bibinfo{author}{\bibfnamefont{M.~E.} \bibnamefont{Schwartz}},
  \bibinfo{author}{\bibfnamefont{M.~Y.} \bibnamefont{Han}},
  \bibinfo{author}{\bibfnamefont{P.}~\bibnamefont{Kim}}, \bibnamefont{and}
  \bibinfo{author}{\bibfnamefont{H.~L.} \bibnamefont{Stormer}},
  \bibinfo{journal}{Phys. Rev. Lett.} \textbf{\bibinfo{volume}{98}},
  \bibinfo{pages}{197403} (\bibinfo{year}{2007}).

\bibitem[{\citenamefont{Deacon et~al.}(2007)\citenamefont{Deacon, Chuang,
  Nicholas, Novoselov, and Geim}}]{DeaconPRB07}
\bibinfo{author}{\bibfnamefont{R.~S.} \bibnamefont{Deacon}},
  \bibinfo{author}{\bibfnamefont{K.-C.} \bibnamefont{Chuang}},
  \bibinfo{author}{\bibfnamefont{R.~J.} \bibnamefont{Nicholas}},
  \bibinfo{author}{\bibfnamefont{K.~S.} \bibnamefont{Novoselov}},
  \bibnamefont{and} \bibinfo{author}{\bibfnamefont{A.~K.} \bibnamefont{Geim}},
  \bibinfo{journal}{Phys. Rev. B} \textbf{\bibinfo{volume}{76}},
  \bibinfo{pages}{081406R} (\bibinfo{year}{2007}).

\bibitem[{\citenamefont{Henriksen et~al.}(2008)\citenamefont{Henriksen, Jiang,
  Tung, Schwartz, Takita, Wang, Kim, and Stormer}}]{HenriksenPRL08}
\bibinfo{author}{\bibfnamefont{E.~A.} \bibnamefont{Henriksen}},
  \bibinfo{author}{\bibfnamefont{Z.}~\bibnamefont{Jiang}},
  \bibinfo{author}{\bibfnamefont{L.-C.} \bibnamefont{Tung}},
  \bibinfo{author}{\bibfnamefont{M.~E.} \bibnamefont{Schwartz}},
  \bibinfo{author}{\bibfnamefont{M.}~\bibnamefont{Takita}},
  \bibinfo{author}{\bibfnamefont{Y.-J.} \bibnamefont{Wang}},
  \bibinfo{author}{\bibfnamefont{P.}~\bibnamefont{Kim}}, \bibnamefont{and}
  \bibinfo{author}{\bibfnamefont{H.~L.} \bibnamefont{Stormer}},
  \bibinfo{journal}{Phys. Rev. Lett.} \textbf{\bibinfo{volume}{100}},
  \bibinfo{pages}{087403} (\bibinfo{year}{2008}).

\bibitem[{\citenamefont{Sadowski et~al.}(2006)\citenamefont{Sadowski, Martinez,
  Potemski, Berger, and de~Heer}}]{SadowskiPRL06}
\bibinfo{author}{\bibfnamefont{M.~L.} \bibnamefont{Sadowski}},
  \bibinfo{author}{\bibfnamefont{G.}~\bibnamefont{Martinez}},
  \bibinfo{author}{\bibfnamefont{M.}~\bibnamefont{Potemski}},
  \bibinfo{author}{\bibfnamefont{C.}~\bibnamefont{Berger}}, \bibnamefont{and}
  \bibinfo{author}{\bibfnamefont{W.~A.} \bibnamefont{de~Heer}},
  \bibinfo{journal}{Phys. Rev. Lett.} \textbf{\bibinfo{volume}{97}},
  \bibinfo{pages}{266405} (\bibinfo{year}{2006}).

\bibitem[{\citenamefont{Sadowski et~al.}(2007)\citenamefont{Sadowski, Martinez,
  Potemski, Berger, and de~Heer}}]{SadowskiSSC07}
\bibinfo{author}{\bibfnamefont{M.~L.} \bibnamefont{Sadowski}},
  \bibinfo{author}{\bibfnamefont{G.}~\bibnamefont{Martinez}},
  \bibinfo{author}{\bibfnamefont{M.}~\bibnamefont{Potemski}},
  \bibinfo{author}{\bibfnamefont{C.}~\bibnamefont{Berger}}, \bibnamefont{and}
  \bibinfo{author}{\bibfnamefont{W.~A.} \bibnamefont{de~Heer}},
  \bibinfo{journal}{Solid State Com.} \textbf{\bibinfo{volume}{143}},
  \bibinfo{pages}{123} (\bibinfo{year}{2007}).

\bibitem[{\citenamefont{Plochocka et~al.}(2008)\citenamefont{Plochocka,
  Faugeras, Orlita, Sadowski, Martinez, Potemski, Goerbig, Fuchs, Berger, and
  de~Heer}}]{PlochockaPRL08}
\bibinfo{author}{\bibfnamefont{P.}~\bibnamefont{Plochocka}},
  \bibinfo{author}{\bibfnamefont{C.}~\bibnamefont{Faugeras}},
  \bibinfo{author}{\bibfnamefont{M.}~\bibnamefont{Orlita}},
  \bibinfo{author}{\bibfnamefont{M.~L.} \bibnamefont{Sadowski}},
  \bibinfo{author}{\bibfnamefont{G.}~\bibnamefont{Martinez}},
  \bibinfo{author}{\bibfnamefont{M.}~\bibnamefont{Potemski}},
  \bibinfo{author}{\bibfnamefont{M.~O.} \bibnamefont{Goerbig}},
  \bibinfo{author}{\bibfnamefont{J.-N.} \bibnamefont{Fuchs}},
  \bibinfo{author}{\bibfnamefont{C.}~\bibnamefont{Berger}}, \bibnamefont{and}
  \bibinfo{author}{\bibfnamefont{W.~A.} \bibnamefont{de~Heer}},
  \bibinfo{journal}{Phys. Rev. Lett.} \textbf{\bibinfo{volume}{100}},
  \bibinfo{pages}{087401} (\bibinfo{year}{2008}).

\bibitem[{\citenamefont{Orlita et~al.}(2008{\natexlab{a}})\citenamefont{Orlita,
  Faugeras, Plochocka, Neugebauer, Martinez, Maude, Barra, Sprinkle, Berger,
  de~Heer et~al.}}]{OrlitaCM08II}
\bibinfo{author}{\bibfnamefont{M.}~\bibnamefont{Orlita}},
  \bibinfo{author}{\bibfnamefont{C.}~\bibnamefont{Faugeras}},
  \bibinfo{author}{\bibfnamefont{P.}~\bibnamefont{Plochocka}},
  \bibinfo{author}{\bibfnamefont{P.}~\bibnamefont{Neugebauer}},
  \bibinfo{author}{\bibfnamefont{G.}~\bibnamefont{Martinez}},
  \bibinfo{author}{\bibfnamefont{D.~K.} \bibnamefont{Maude}},
  \bibinfo{author}{\bibfnamefont{A.-L.} \bibnamefont{Barra}},
  \bibinfo{author}{\bibfnamefont{M.}~\bibnamefont{Sprinkle}},
  \bibinfo{author}{\bibfnamefont{C.}~\bibnamefont{Berger}},
  \bibinfo{author}{\bibfnamefont{W.~A.} \bibnamefont{de~Heer}},
  \bibnamefont{et~al.}, \emph{\bibinfo{title}{arxiv:0808.3662}}
  (\bibinfo{year}{2008}{\natexlab{a}}).

\bibitem[{\citenamefont{Li et~al.}(2006)\citenamefont{Li, Tsai, Padilla,
  Dordevic, Burch, Wang, and Basov}}]{LiPRB06}
\bibinfo{author}{\bibfnamefont{Z.~Q.} \bibnamefont{Li}},
  \bibinfo{author}{\bibfnamefont{S.-W.} \bibnamefont{Tsai}},
  \bibinfo{author}{\bibfnamefont{W.~J.} \bibnamefont{Padilla}},
  \bibinfo{author}{\bibfnamefont{S.~V.} \bibnamefont{Dordevic}},
  \bibinfo{author}{\bibfnamefont{K.~S.} \bibnamefont{Burch}},
  \bibinfo{author}{\bibfnamefont{Y.~J.} \bibnamefont{Wang}}, \bibnamefont{and}
  \bibinfo{author}{\bibfnamefont{D.~N.} \bibnamefont{Basov}},
  \bibinfo{journal}{Phys. Rev. B} \textbf{\bibinfo{volume}{74}},
  \bibinfo{pages}{195404} (\bibinfo{year}{2006}).

\bibitem[{\citenamefont{Orlita et~al.}(2008{\natexlab{b}})\citenamefont{Orlita,
  Faugeras, Martinez, Maude, Sadowski, and Potemski}}]{OrlitaPRL08}
\bibinfo{author}{\bibfnamefont{M.}~\bibnamefont{Orlita}},
  \bibinfo{author}{\bibfnamefont{C.}~\bibnamefont{Faugeras}},
  \bibinfo{author}{\bibfnamefont{G.}~\bibnamefont{Martinez}},
  \bibinfo{author}{\bibfnamefont{D.~K.} \bibnamefont{Maude}},
  \bibinfo{author}{\bibfnamefont{M.~L.} \bibnamefont{Sadowski}},
  \bibnamefont{and} \bibinfo{author}{\bibfnamefont{M.}~\bibnamefont{Potemski}},
  \bibinfo{journal}{Phys. Rev. Lett.} \textbf{\bibinfo{volume}{100}},
  \bibinfo{pages}{136403} (\bibinfo{year}{2008}{\natexlab{b}}).

\bibitem[{\citenamefont{Orlita et~al.}(2008{\natexlab{c}})\citenamefont{Orlita,
  Faugeras, Martinez, Maude, Sadowski, Schneider, and Potemski}}]{OrlitaJPCM08}
\bibinfo{author}{\bibfnamefont{M.}~\bibnamefont{Orlita}},
  \bibinfo{author}{\bibfnamefont{C.}~\bibnamefont{Faugeras}},
  \bibinfo{author}{\bibfnamefont{G.}~\bibnamefont{Martinez}},
  \bibinfo{author}{\bibfnamefont{D.~K.} \bibnamefont{Maude}},
  \bibinfo{author}{\bibfnamefont{M.~L.} \bibnamefont{Sadowski}},
  \bibinfo{author}{\bibfnamefont{J.~M.} \bibnamefont{Schneider}},
  \bibnamefont{and} \bibinfo{author}{\bibfnamefont{M.}~\bibnamefont{Potemski}},
  \bibinfo{journal}{Journal of Physics: Condensed Matter}
  \textbf{\bibinfo{volume}{20}}, \bibinfo{pages}{454223}
  (\bibinfo{year}{2008}{\natexlab{c}}).

\bibitem[{\citenamefont{Garcia-Flores et~al.}(2008)\citenamefont{Garcia-Flores,
  Terashita, Granado, and Kopelevich}}]{Garcia-FloresCM08}
\bibinfo{author}{\bibfnamefont{A.~F.} \bibnamefont{Garcia-Flores}},
  \bibinfo{author}{\bibfnamefont{H.}~\bibnamefont{Terashita}},
  \bibinfo{author}{\bibfnamefont{E.}~\bibnamefont{Granado}}, \bibnamefont{and}
  \bibinfo{author}{\bibfnamefont{Y.}~\bibnamefont{Kopelevich}},
  \emph{\bibinfo{title}{arxiv:0807.1343}} (\bibinfo{year}{2008}).

\bibitem[{\citenamefont{Iyengar et~al.}(2007)\citenamefont{Iyengar, Wang,
  Fertig, and Brey}}]{IyengarPRB07}
\bibinfo{author}{\bibfnamefont{A.}~\bibnamefont{Iyengar}},
  \bibinfo{author}{\bibfnamefont{J.}~\bibnamefont{Wang}},
  \bibinfo{author}{\bibfnamefont{H.~A.} \bibnamefont{Fertig}},
  \bibnamefont{and} \bibinfo{author}{\bibfnamefont{L.}~\bibnamefont{Brey}},
  \bibinfo{journal}{Phys. Rev. B} \textbf{\bibinfo{volume}{75}},
  \bibinfo{pages}{125430} (\bibinfo{year}{2007}).

\bibitem[{\citenamefont{Gusynin et~al.}(2007)\citenamefont{Gusynin, Sharapov,
  and Carbotte}}]{GusyninPRL07}
\bibinfo{author}{\bibfnamefont{V.~P.} \bibnamefont{Gusynin}},
  \bibinfo{author}{\bibfnamefont{S.~G.} \bibnamefont{Sharapov}},
  \bibnamefont{and} \bibinfo{author}{\bibfnamefont{J.~P.}
  \bibnamefont{Carbotte}}, \bibinfo{journal}{Phys. Rev. Lett.}
  \textbf{\bibinfo{volume}{98}}, \bibinfo{pages}{157402}
  (\bibinfo{year}{2007}).

\bibitem[{\citenamefont{Abergel and Fal'ko}(2007)}]{AbergelPRB07}
\bibinfo{author}{\bibfnamefont{D.~S.~L.} \bibnamefont{Abergel}}
  \bibnamefont{and} \bibinfo{author}{\bibfnamefont{V.~I.}
  \bibnamefont{Fal'ko}}, \bibinfo{journal}{Phys. Rev. B}
  \textbf{\bibinfo{volume}{75}}, \bibinfo{pages}{155430}
  (\bibinfo{year}{2007}).

\bibitem[{\citenamefont{Koshino and Ando}(2008)}]{KoshinoPRB08}
\bibinfo{author}{\bibfnamefont{M.}~\bibnamefont{Koshino}} \bibnamefont{and}
  \bibinfo{author}{\bibfnamefont{T.}~\bibnamefont{Ando}},
  \bibinfo{journal}{Phys. Rev. B} \textbf{\bibinfo{volume}{77}},
  \bibinfo{pages}{115313} (\bibinfo{year}{2008}).

\bibitem[{\citenamefont{Bychkov and Martinez}(2008)}]{BychkovPRB08}
\bibinfo{author}{\bibfnamefont{Y.~A.} \bibnamefont{Bychkov}} \bibnamefont{and}
  \bibinfo{author}{\bibfnamefont{G.}~\bibnamefont{Martinez}},
  \bibinfo{journal}{Phys. Rev. B} \textbf{\bibinfo{volume}{77}},
  \bibinfo{pages}{125417} (\bibinfo{year}{2008}).

\bibitem[{\citenamefont{Nair et~al.}(2008)\citenamefont{Nair, Blake,
  Grigorenko, Novoselov, Booth, Stauber, Peres, and Geim}}]{NairScience08}
\bibinfo{author}{\bibfnamefont{R.~R.} \bibnamefont{Nair}},
  \bibinfo{author}{\bibfnamefont{P.}~\bibnamefont{Blake}},
  \bibinfo{author}{\bibfnamefont{A.~N.} \bibnamefont{Grigorenko}},
  \bibinfo{author}{\bibfnamefont{K.~S.} \bibnamefont{Novoselov}},
  \bibinfo{author}{\bibfnamefont{T.~J.} \bibnamefont{Booth}},
  \bibinfo{author}{\bibfnamefont{T.}~\bibnamefont{Stauber}},
  \bibinfo{author}{\bibfnamefont{N.~M.~R.} \bibnamefont{Peres}},
  \bibnamefont{and} \bibinfo{author}{\bibfnamefont{A.~K.} \bibnamefont{Geim}},
  \bibinfo{journal}{Science} \textbf{\bibinfo{volume}{320}},
  \bibinfo{pages}{1308} (\bibinfo{year}{2008}).

\bibitem[{\citenamefont{Kuzmenko
  et~al.}(2008{\natexlab{a}})\citenamefont{Kuzmenko, van Heumen, Carbone, and
  van~der Marel}}]{KuzmenkoPRL08}
\bibinfo{author}{\bibfnamefont{A.~B.} \bibnamefont{Kuzmenko}},
  \bibinfo{author}{\bibfnamefont{E.}~\bibnamefont{van Heumen}},
  \bibinfo{author}{\bibfnamefont{F.}~\bibnamefont{Carbone}}, \bibnamefont{and}
  \bibinfo{author}{\bibfnamefont{D.}~\bibnamefont{van~der Marel}},
  \bibinfo{journal}{Phys. Rev. Lett.} \textbf{\bibinfo{volume}{100}},
  \bibinfo{pages}{117401} (\bibinfo{year}{2008}{\natexlab{a}}).

\bibitem[{\citenamefont{Wang et~al.}(2008)\citenamefont{Wang, Zhang, Tian,
  Girit, Zettl, Crommie, and Shen}}]{WangScience08}
\bibinfo{author}{\bibfnamefont{F.}~\bibnamefont{Wang}},
  \bibinfo{author}{\bibfnamefont{Y.}~\bibnamefont{Zhang}},
  \bibinfo{author}{\bibfnamefont{C.}~\bibnamefont{Tian}},
  \bibinfo{author}{\bibfnamefont{C.}~\bibnamefont{Girit}},
  \bibinfo{author}{\bibfnamefont{A.}~\bibnamefont{Zettl}},
  \bibinfo{author}{\bibfnamefont{M.}~\bibnamefont{Crommie}}, \bibnamefont{and}
  \bibinfo{author}{\bibfnamefont{Y.~R.} \bibnamefont{Shen}},
  \bibinfo{journal}{Science} \textbf{\bibinfo{volume}{320}},
  \bibinfo{pages}{206} (\bibinfo{year}{2008}).

\bibitem[{\citenamefont{Li et~al.}(2008{\natexlab{a}})\citenamefont{Li,
  Henriksen, Jiang, Hao, Martin, Kim, Stormer, and Basov}}]{LiNaturePhys08}
\bibinfo{author}{\bibfnamefont{Z.~Q.} \bibnamefont{Li}},
  \bibinfo{author}{\bibfnamefont{E.~A.} \bibnamefont{Henriksen}},
  \bibinfo{author}{\bibfnamefont{Z.}~\bibnamefont{Jiang}},
  \bibinfo{author}{\bibfnamefont{Z.}~\bibnamefont{Hao}},
  \bibinfo{author}{\bibfnamefont{M.~C.} \bibnamefont{Martin}},
  \bibinfo{author}{\bibfnamefont{P.}~\bibnamefont{Kim}},
  \bibinfo{author}{\bibfnamefont{H.~L.} \bibnamefont{Stormer}},
  \bibnamefont{and} \bibinfo{author}{\bibfnamefont{D.~N.} \bibnamefont{Basov}},
  \bibinfo{journal}{Nature Physics} \textbf{\bibinfo{volume}{4}},
  \bibinfo{pages}{532} (\bibinfo{year}{2008}{\natexlab{a}}).

\bibitem[{\citenamefont{Kuzmenko
  et~al.}(2008{\natexlab{b}})\citenamefont{Kuzmenko, van Heumen, van~der Marel,
  Lerch, Blake, Novoselov, and Geim}}]{KuzmenkoCM08}
\bibinfo{author}{\bibfnamefont{A.~B.} \bibnamefont{Kuzmenko}},
  \bibinfo{author}{\bibfnamefont{E.}~\bibnamefont{van Heumen}},
  \bibinfo{author}{\bibfnamefont{D.}~\bibnamefont{van~der Marel}},
  \bibinfo{author}{\bibfnamefont{P.}~\bibnamefont{Lerch}},
  \bibinfo{author}{\bibfnamefont{P.}~\bibnamefont{Blake}},
  \bibinfo{author}{\bibfnamefont{K.~S.} \bibnamefont{Novoselov}},
  \bibnamefont{and} \bibinfo{author}{\bibfnamefont{A.~K.} \bibnamefont{Geim}},
  \emph{\bibinfo{title}{arxiv:0810.2400}} (\bibinfo{year}{2008}{\natexlab{b}}).

\bibitem[{\citenamefont{Zhang et~al.}(2008)\citenamefont{Zhang, Li, Basov,
  Fogler, Hao, and Martin}}]{ZhangCM08}
\bibinfo{author}{\bibfnamefont{L.~M.} \bibnamefont{Zhang}},
  \bibinfo{author}{\bibfnamefont{Z.~Q.} \bibnamefont{Li}},
  \bibinfo{author}{\bibfnamefont{D.~N.} \bibnamefont{Basov}},
  \bibinfo{author}{\bibfnamefont{M.~M.} \bibnamefont{Fogler}},
  \bibinfo{author}{\bibfnamefont{Z.}~\bibnamefont{Hao}}, \bibnamefont{and}
  \bibinfo{author}{\bibfnamefont{M.~C.} \bibnamefont{Martin}},
  \emph{\bibinfo{title}{arxiv:0809.1898}} (\bibinfo{year}{2008}).

\bibitem[{\citenamefont{McCann et~al.}(2007)\citenamefont{McCann, Abergel, and
  Fal'ko}}]{McCannSSC07}
\bibinfo{author}{\bibfnamefont{E.}~\bibnamefont{McCann}},
  \bibinfo{author}{\bibfnamefont{D.~S.~L.} \bibnamefont{Abergel}},
  \bibnamefont{and} \bibinfo{author}{\bibfnamefont{V.~I.}
  \bibnamefont{Fal'ko}}, \bibinfo{journal}{Solid state communicatins}
  \textbf{\bibinfo{volume}{143}}, \bibinfo{pages}{110} (\bibinfo{year}{2007}).

\bibitem[{\citenamefont{Stauber et~al.}(2008)\citenamefont{Stauber, Peres, and
  Geim}}]{StauberPRB08}
\bibinfo{author}{\bibfnamefont{T.}~\bibnamefont{Stauber}},
  \bibinfo{author}{\bibfnamefont{N.~M.~R.} \bibnamefont{Peres}},
  \bibnamefont{and} \bibinfo{author}{\bibfnamefont{A.~K.} \bibnamefont{Geim}},
  \bibinfo{journal}{Phys. Rev. B} \textbf{\bibinfo{volume}{78}},
  \bibinfo{pages}{085432} (\bibinfo{year}{2008}).

\bibitem[{\citenamefont{Nicol and Carbotte}(2008)}]{NicolPRB08}
\bibinfo{author}{\bibfnamefont{E.~J.} \bibnamefont{Nicol}} \bibnamefont{and}
  \bibinfo{author}{\bibfnamefont{J.~P.} \bibnamefont{Carbotte}},
  \bibinfo{journal}{Phys. Rev. B} \textbf{\bibinfo{volume}{77}},
  \bibinfo{pages}{155409} (\bibinfo{year}{2008}).

\bibitem[{\citenamefont{Dawlaty et~al.}(2008)\citenamefont{Dawlaty, Shivaraman,
  Chandrashekhar, Rana, and Spencer}}]{DawlatyAPL08}
\bibinfo{author}{\bibfnamefont{J.~M.} \bibnamefont{Dawlaty}},
  \bibinfo{author}{\bibfnamefont{S.}~\bibnamefont{Shivaraman}},
  \bibinfo{author}{\bibfnamefont{M.}~\bibnamefont{Chandrashekhar}},
  \bibinfo{author}{\bibfnamefont{F.}~\bibnamefont{Rana}}, \bibnamefont{and}
  \bibinfo{author}{\bibfnamefont{M.~G.} \bibnamefont{Spencer}},
  \bibinfo{journal}{Appl. Phys. Lett.} \textbf{\bibinfo{volume}{92}},
  \bibinfo{pages}{042116} (\bibinfo{year}{2008}).

\bibitem[{\citenamefont{Sun et~al.}(2008)\citenamefont{Sun, Wu, Divin, Li,
  Berger, de~Heer, First, and Norris}}]{SunPRL08}
\bibinfo{author}{\bibfnamefont{D.}~\bibnamefont{Sun}},
  \bibinfo{author}{\bibfnamefont{Z.-K.} \bibnamefont{Wu}},
  \bibinfo{author}{\bibfnamefont{C.}~\bibnamefont{Divin}},
  \bibinfo{author}{\bibfnamefont{X.}~\bibnamefont{Li}},
  \bibinfo{author}{\bibfnamefont{C.}~\bibnamefont{Berger}},
  \bibinfo{author}{\bibfnamefont{W.~A.} \bibnamefont{de~Heer}},
  \bibinfo{author}{\bibfnamefont{P.~N.} \bibnamefont{First}}, \bibnamefont{and}
  \bibinfo{author}{\bibfnamefont{T.~B.} \bibnamefont{Norris}},
  \bibinfo{journal}{Phys. Rev. Lett.} \textbf{\bibinfo{volume}{101}},
  \bibinfo{pages}{157402} (\bibinfo{year}{2008}).

\bibitem[{\citenamefont{Berger et~al.}(2004)\citenamefont{Berger, Song, Li, Li,
  Ogbazghi, Feng, Dai, Marchenkov, Conrad, First et~al.}}]{BergerJPCB04}
\bibinfo{author}{\bibfnamefont{C.}~\bibnamefont{Berger}},
  \bibinfo{author}{\bibfnamefont{Z.}~\bibnamefont{Song}},
  \bibinfo{author}{\bibfnamefont{T.}~\bibnamefont{Li}},
  \bibinfo{author}{\bibfnamefont{X.}~\bibnamefont{Li}},
  \bibinfo{author}{\bibfnamefont{A.~Y.} \bibnamefont{Ogbazghi}},
  \bibinfo{author}{\bibfnamefont{R.}~\bibnamefont{Feng}},
  \bibinfo{author}{\bibfnamefont{Z.}~\bibnamefont{Dai}},
  \bibinfo{author}{\bibfnamefont{A.~N.} \bibnamefont{Marchenkov}},
  \bibinfo{author}{\bibfnamefont{E.~H.} \bibnamefont{Conrad}},
  \bibinfo{author}{\bibfnamefont{P.~N.} \bibnamefont{First}},
  \bibnamefont{et~al.}, \bibinfo{journal}{J. Phys. Chem. B}
  \textbf{\bibinfo{volume}{108}}, \bibinfo{pages}{19912}
  (\bibinfo{year}{2004}).

\bibitem[{\citenamefont{Berger et~al.}(2006)\citenamefont{Berger, Song, Li, Wu,
  Brown, Naud, Mayou, Li, Hass, Marchenkov et~al.}}]{BergerScience06}
\bibinfo{author}{\bibfnamefont{C.}~\bibnamefont{Berger}},
  \bibinfo{author}{\bibfnamefont{Z.}~\bibnamefont{Song}},
  \bibinfo{author}{\bibfnamefont{X.}~\bibnamefont{Li}},
  \bibinfo{author}{\bibfnamefont{X.}~\bibnamefont{Wu}},
  \bibinfo{author}{\bibfnamefont{N.}~\bibnamefont{Brown}},
  \bibinfo{author}{\bibfnamefont{C.}~\bibnamefont{Naud}},
  \bibinfo{author}{\bibfnamefont{D.}~\bibnamefont{Mayou}},
  \bibinfo{author}{\bibfnamefont{T.}~\bibnamefont{Li}},
  \bibinfo{author}{\bibfnamefont{J.}~\bibnamefont{Hass}},
  \bibinfo{author}{\bibfnamefont{A.~N.} \bibnamefont{Marchenkov}},
  \bibnamefont{et~al.}, \bibinfo{journal}{Science}
  \textbf{\bibinfo{volume}{312}}, \bibinfo{pages}{1191} (\bibinfo{year}{2006}).

\bibitem[{\citenamefont{Ferrari et~al.}(2006)\citenamefont{Ferrari, Meyer,
  Scardaci, Casiraghi, Lazzeri, Mauri, Piscanec, Jiang, Novoselov, Roth
  et~al.}}]{FerrariPRL06}
\bibinfo{author}{\bibfnamefont{A.~C.} \bibnamefont{Ferrari}},
  \bibinfo{author}{\bibfnamefont{J.~C.} \bibnamefont{Meyer}},
  \bibinfo{author}{\bibfnamefont{V.}~\bibnamefont{Scardaci}},
  \bibinfo{author}{\bibfnamefont{C.}~\bibnamefont{Casiraghi}},
  \bibinfo{author}{\bibfnamefont{M.}~\bibnamefont{Lazzeri}},
  \bibinfo{author}{\bibfnamefont{F.}~\bibnamefont{Mauri}},
  \bibinfo{author}{\bibfnamefont{S.}~\bibnamefont{Piscanec}},
  \bibinfo{author}{\bibfnamefont{D.}~\bibnamefont{Jiang}},
  \bibinfo{author}{\bibfnamefont{K.~S.} \bibnamefont{Novoselov}},
  \bibinfo{author}{\bibfnamefont{S.}~\bibnamefont{Roth}}, \bibnamefont{et~al.},
  \bibinfo{journal}{Phys. Rev. Lett.} \textbf{\bibinfo{volume}{97}},
  \bibinfo{pages}{187401} (\bibinfo{year}{2006}).

\bibitem[{\citenamefont{Faugeras et~al.}(2008)\citenamefont{Faugeras,
  Nerri\`{e}re, Potemski, Mahmood, Dujardin, Berger, and
  de~Heer}}]{FaugerasAPL08}
\bibinfo{author}{\bibfnamefont{C.}~\bibnamefont{Faugeras}},
  \bibinfo{author}{\bibfnamefont{A.}~\bibnamefont{Nerri\`{e}re}},
  \bibinfo{author}{\bibfnamefont{M.}~\bibnamefont{Potemski}},
  \bibinfo{author}{\bibfnamefont{A.}~\bibnamefont{Mahmood}},
  \bibinfo{author}{\bibfnamefont{E.}~\bibnamefont{Dujardin}},
  \bibinfo{author}{\bibfnamefont{C.}~\bibnamefont{Berger}}, \bibnamefont{and}
  \bibinfo{author}{\bibfnamefont{W.~A.} \bibnamefont{de~Heer}},
  \bibinfo{journal}{Appl. Phys. Lett.} \textbf{\bibinfo{volume}{92}},
  \bibinfo{eid}{011914} (\bibinfo{year}{2008}).

\bibitem[{\citenamefont{Hass et~al.}(2008)\citenamefont{Hass, Varchon,
  Mill\'{a}n-Otoya, Sprinkle, Sharma, de~Heer, Berger, First, Magaud, and
  Conrad}}]{HassPRL08}
\bibinfo{author}{\bibfnamefont{J.}~\bibnamefont{Hass}},
  \bibinfo{author}{\bibfnamefont{F.}~\bibnamefont{Varchon}},
  \bibinfo{author}{\bibfnamefont{J.~E.} \bibnamefont{Mill\'{a}n-Otoya}},
  \bibinfo{author}{\bibfnamefont{M.}~\bibnamefont{Sprinkle}},
  \bibinfo{author}{\bibfnamefont{N.}~\bibnamefont{Sharma}},
  \bibinfo{author}{\bibfnamefont{W.~A.} \bibnamefont{de~Heer}},
  \bibinfo{author}{\bibfnamefont{C.}~\bibnamefont{Berger}},
  \bibinfo{author}{\bibfnamefont{P.~N.} \bibnamefont{First}},
  \bibinfo{author}{\bibfnamefont{L.}~\bibnamefont{Magaud}}, \bibnamefont{and}
  \bibinfo{author}{\bibfnamefont{E.~H.} \bibnamefont{Conrad}},
  \bibinfo{journal}{Phys. Rev. Lett.} \textbf{\bibinfo{volume}{100}},
  \bibinfo{pages}{125504} (\bibinfo{year}{2008}).

\bibitem[{\citenamefont{Slonczewski and Weiss}(1958)}]{SlonczewskiPR58}
\bibinfo{author}{\bibfnamefont{J.~C.} \bibnamefont{Slonczewski}}
  \bibnamefont{and} \bibinfo{author}{\bibfnamefont{P.~R.} \bibnamefont{Weiss}},
  \bibinfo{journal}{Phys. Rev.} \textbf{\bibinfo{volume}{109}},
  \bibinfo{pages}{272} (\bibinfo{year}{1958}).

\bibitem[{\citenamefont{McClure}(1957)}]{McClurePR57}
\bibinfo{author}{\bibfnamefont{J.~W.} \bibnamefont{McClure}},
  \bibinfo{journal}{Phys. Rev.} \textbf{\bibinfo{volume}{108}},
  \bibinfo{pages}{606} (\bibinfo{year}{1957}).

\bibitem[{\citenamefont{McClure}(1960)}]{McClurePR60}
\bibinfo{author}{\bibfnamefont{J.~W.} \bibnamefont{McClure}},
  \bibinfo{journal}{Phys. Rev.} \textbf{\bibinfo{volume}{119}},
  \bibinfo{pages}{612} (\bibinfo{year}{1960}).

\bibitem[{\citenamefont{Toy et~al.}(1977)\citenamefont{Toy, Dresselhaus, and
  Dresselhaus}}]{ToyPRB77}
\bibinfo{author}{\bibfnamefont{W.~W.} \bibnamefont{Toy}},
  \bibinfo{author}{\bibfnamefont{M.~S.} \bibnamefont{Dresselhaus}},
  \bibnamefont{and}
  \bibinfo{author}{\bibfnamefont{G.}~\bibnamefont{Dresselhaus}},
  \bibinfo{journal}{Phys. Rev. B} \textbf{\bibinfo{volume}{15}},
  \bibinfo{pages}{4077} (\bibinfo{year}{1977}).

\bibitem[{\citenamefont{Gr{\"{u}}neis et~al.}(2008)\citenamefont{Gr{\"{u}}neis,
  Attaccalite, Wirtz, Shiozawa, Saito, Pichler, and Rubio}}]{GruneisCM08}
\bibinfo{author}{\bibfnamefont{A.}~\bibnamefont{Gr{\"{u}}neis}},
  \bibinfo{author}{\bibfnamefont{C.}~\bibnamefont{Attaccalite}},
  \bibinfo{author}{\bibfnamefont{L.}~\bibnamefont{Wirtz}},
  \bibinfo{author}{\bibfnamefont{H.}~\bibnamefont{Shiozawa}},
  \bibinfo{author}{\bibfnamefont{R.}~\bibnamefont{Saito}},
  \bibinfo{author}{\bibfnamefont{T.}~\bibnamefont{Pichler}}, \bibnamefont{and}
  \bibinfo{author}{\bibfnamefont{A.}~\bibnamefont{Rubio}},
  \emph{\bibinfo{title}{0808.1467}} (\bibinfo{year}{2008}).

\bibitem[{\citenamefont{Li et~al.}(2008{\natexlab{b}})\citenamefont{Li, Luican,
  and Andrei}}]{LiCM08}
\bibinfo{author}{\bibfnamefont{G.}~\bibnamefont{Li}},
  \bibinfo{author}{\bibfnamefont{A.}~\bibnamefont{Luican}}, \bibnamefont{and}
  \bibinfo{author}{\bibfnamefont{E.~Y.} \bibnamefont{Andrei}},
  \emph{\bibinfo{title}{arxiv:0803.4016}} (\bibinfo{year}{2008}{\natexlab{b}}).

\end{thebibliography}

\end{document}